\documentstyle[aps,epsf,multicol]{revtex}

\begin{document}
\title{
Stability of trapped Bose-Einstein 
condensates~\thanks{To appear in Physical Review A}}
\author{F.Kh. Abdullaev$^{1}$\footnote{Permanent address:
Physical-Technical Institute, Tashkent, Uzbekistan},
A. Gammal$^{1}$, Lauro Tomio$^{1}$, 
and T. Frederico$^{2}$
}
\address{
$^{1}$Instituto de F\'{\i}sica Te\'{o}rica,
Universidade Estadual Paulista, 
01405-900 S\~{a}o Paulo, Brazil \\
$^{2}$Departamento de F\'{\i }sica, Instituto Tecnol\'{o}gico da
Aeron\'{a}utica, \\
Centro T\'{e}cnico Aeroespacial, 12228-900 S\~{a}o Jos\'{e} dos Campos,
SP, Brazil
}
\date{\today}
\maketitle
\begin{abstract}
In three-dimensional trapped Bose-Einstein condensate (BEC), 
described by the time-dependent Gross-Pitaevskii-Ginzburg
equation, we study the effect of initial conditions on
stability using a Gaussian variational approach and exact 
numerical simulations. We also discuss the validity of the
criterion for stability suggested by Vakhitov and Kolokolov. 
The maximum initial chirp (initial focusing defocusing of
cloud) that can lead a stable condensate to  collapse even
before the number of atoms reaches its critical limit
is obtained for several specific cases. 
When we consider two- and three-body nonlinear terms, with 
negative cubic and positive quintic terms, we have the
conditions for the existence of two phases in the condensate. 
In this case, the magnitude of the oscillations between the two
phases are studied considering sufficient large initial chirps.
The occurrence of collapse in a BEC with repulsive two-body
interaction is also shown to be possible.
\end{abstract} 
\begin{multicols}{2}
\section{Introduction}
In trapped Bose-Einstein condensates(BEC), it is well-known the
occurrence of collapse when the two-body interaction is attractive and if
the number of atoms ${N}$ exceeds a critical value $N_c$, as in the case
of atomic condensates with $^{7}\mbox{Li}$~\cite{pita}.
In this case, experiments with attractive two-body interaction
have been performed~\cite{hulet}, with results consistent with the
limitation in the number of atoms and with the grow and collapse
scenario. As the non-linear terms presented in the non-linear
Schr\"odinger equation (NLSE) (cubic, quintic, and so on) are due to
the first terms of an expansion of the effective many-body
interaction in the mean-field approximation~\cite{jpb00}, a study of
stability of the equation that describes BEC is necessary,
considering the observed  few-body interactions that occur in the
atomic gas. When the overall sign of the effective many-body
interaction is negative, we have the conditions for the occurrence of
collapse, with the stability of the condensate being restricted by
the number of atoms.

We should stress that the criteria for stability usually  involves one
parameter: the number of atoms. Here, in order to introduce a criterion
for stability of a trapped condensate with two and three-body effective
interactions, we introduce another parameter: a chirp, that is related
to the initial focusing (defocusing) of the cloud.  Without trap, a 
chirp parameter was previously introduced in Ref.~\cite{andertal}
to study  the stability of the NLSE; and with trap, the stability
has been studied in Refs.~\cite{fetter,perez,stoof}. One should also
note that an oscillating condensate always has a chirped wave
function, where the chirp is proportional to the time variation of
the width, which is a periodic function of time. Thus, a real
condensate, for example in optical traps, always has nonzero initial
chirp, that should be taken into account.

Collapses for two and three dimension ($2D$ and $3D$) geometries of the
NLSE with trap potential has been investigated in
Refs.~\cite{pita,Wadati}, where the corresponding critical numbers
were obtained \cite{Ruprecht,pre99}.  These works are extensions of the
Zakharov and co-workers theory of collapse
\cite{Zakharov,BergeR,ZakharovR,weins} to
the case when an additional trap potential is included.  
The trap potential helps to prevent the collapsing process, which 
cannot be avoided if the number of atoms exceeds the critical limit. 
In Refs.~\cite{Wadati,Berge}, considering the moments method in the
analysis of the equation for the average squared value of the width,
$\langle a^2 \rangle$, it was emphasized the influence of
the corresponding initial condition on the collapse.

Variational approaches have also been used by many authors for the 
analysis of stability of NLSE, and proved to be useful in many aspects
(see Ref.~\cite{andertal}, for example).
The variational approach (VA), using a simple Gaussian ansatz, gives a
reasonable description of the conditions for the collapse and also an
approximate value for the critical number of atoms. It is interesting to
apply the time dependent variational approach to describe the dynamics of
BEC with attractive two-body interactions
and in particular to study the influence of the initial conditions (using
a chirp constant, $b$, in the wave-function) on the stability of BEC. 
This can be important for the condensation in optical traps and in
strongly inhomogeneous traps.
Also recently, in Ref.~\cite{Huepe}, the stability analysis
was performed, using both VA and exact calculations, considering that the
condensate can unstabilize by tunneling, due to quantum fluctuations.

When the two-body interaction term is switched off, for example by 
tunning an external magnetic field~\cite{avaria}, 
the three-body interaction term can play an important role. 
Thus, it is interesting to study the collapse conditions in the model
with trap potential and an attractive three-body interaction 
(quintic) term. Recently, in Ref.~\cite{Gaididei2}, it was investigated
the collapse in one dimensional ($1D$) model with attractive three-body
interaction. The trap was narrow and induced the radiative loss of atoms,
reducing the number below the critical value, which made an
arrest of the collapse possible. The two-dimensional ($2D$) case, with
cubic
nonlinearity (two-body interactions), may be useful in the
modulation theory around the Townes soliton~\cite{Fibich}. Then it is
possible to estimate the role played by the continuum. 

In the present work, we first consider a review of the stability 
criteria, without application of initial chirp. Next, we analyze the 
role of an initial chirp in the wave-function of stable  condensate
with attractive two-body nonlinear term, which can be useful for actual
experimental analysis in BEC. A generalization of this study of stability
is considered with the inclusion of three-body interaction term in the
NLSE.
Such study can be relevant in the perspective of atomic systems with
enhanced three-body effects, that can happen as the two-body scattering
length is altered~\cite{avaria}.

When a positive three-body interaction term is included in a trapped
three-dimensional ($3D$) NLSE with negative two-body interaction, 
it is already known the possibility of extending the region of stability
of the equation to larger number of atoms, with the occurrence of two
stable phases in the condensate~\cite{Akhmediev,Gammal}. In
Ref.~\cite{jpb00}, it was calculated the frequency of the collective
excitations in denser (liquid) and dilute (gas) phases of the condensate.
Here, we also investigate a bifurcation phenomena related with the time
variation of the width parameter of the theory, by switching the
oscillations from one phase to the other. The present study is done with
the aid of a time-dependent variational approach for the collapse of the
atomic cloud in $3D$ model with harmonic trap potential. We compare the
predictions of the VA with exact numerical simulations of partial
differential equation (PDE). As detailed in our conclusions, the Gaussian
VA gives a good estimate for the observables in the region of stability,
and it starts to deviate from the exact results when the system is close
to the collapse conditions and particularly for the unstable solutions
(maxima) of the total energy. For this regions, where the Gaussian VA
fails, one should improve the ansatz or take the results as a qualitative
picture to guide the exact numerical calculations.

In cases without traps, the NLSE  was previously studied by
Vakhitov and Kolokolov (VK)~\cite{VK}, where a criterion for
stability was settled. This and other criteria for stability
in nonlinear systems have been studied and extensively used by
many authors (see \cite{bara,BergeR} and references therein). The VK
criterion was recently detailed in Ref.~\cite{BergeR}.
When a trap potential was added to a NLSE with a negative cubic term,
an alternative stability criterion was derived in
Ref.~\cite{stoof}, after considering similarities between BEC 
atomic systems and compact object like neutron stars. 
Recently, the VK criterion for trapped BEC with cubic term was
formally demonstrated in Ref.~\cite{Kivshar}.

In recent numerical studies~\cite{jpb00,Gammal,Akhmediev},
one can also observe that the VK criterion is not
generally valid; it cannot be extended, for example, to the case of a
trapped system with both attractive cubic and repulsive quintic
terms. The solutions can be stable irrespectively to the sign of the
derivative of the eigenvalues of the NLSE with respect to the number
of atoms, $d\mu/dN$. When $g_3$ is positive, and for large enough
number os atoms, the NLSE is stable and $d\mu/dN$ can be positive
(see, for example, Ref.~\cite{jpb00}), implying that the criterion
cannot be extended to trapped NLSE with two and three-body terms.
Here, the nonvalidity of the VK criterion is clearly verified in case 
we have a positive harmonic trap, with two and three-body nonlinear
terms with opposite signs. 

We extend the study of the stability of the NLSE with cubic (two-body)
and quintic (three-body) terms, through the analysis of chirp response,
and through the frequency of collective excitations. 
By applying the perturbation techniques to the theory of 
nonlinear oscillations~\cite{Landau}, we also derive the
frequency of weakly nonlinear collective excitations.
This study extend numerical calculations performed in
Ref.~\cite{jpb00} for the frequencies of collective excitations.

One should also note that such study is of actual interest, as
recently in Ref.~\cite{Hec} is was reported observations of nonlinear
oscillations in BEC of a gas with rubidium atoms.
The present study can be of interest not only to atomic BEC but also
to other branches of physics and mathematics, where non-linear
effective interactions are added to a trapped potential in the time
dependent Schr\"odinger equation, such as in optics and soliton
physics.

In Section II, it is given a description of the model, using a
variational approach for a system with $D$ dimensions.
In Section III, we derive an expression for the chirp parameter,
that is considered in our analysis of stability. The maximum
initial chirp to keep the system oscillating in the same phase,
around a minimum of the energy, without collapsing or without
phase-transition, is obtained in this section by considering both
variational and exact numerical calculations. 
In Section IV, we follow the study of Section III, and also
obtain the frequencies for the linear and nonlinear oscillations.
In this section we consider the presence of both, cubic and quintic
terms, in several different configurations. The VK criterion of
stability is discussed in section V.
Finally, in Section VI, we present our concluding remarks.

\section{Description of the model}

In this section we will consider  the dynamics described by the NLSE with
harmonic potential and with cubic and quintic terms. Using a variational
approach for a system with $D$ dimensions, we derive from the Lagrangian
the expression for the anharmonic potential considering the
mechanical analogy.

The time-dependent Gross-Pitaevskii-Ginzburg (GPG) equation that will be  
considered in the present approach, is given by 
\begin{equation}
2{\rm i}\psi_t = -\Delta \psi + V\psi + \lambda_2
|\psi|^2 \psi + \lambda_3 |\psi|^4 \psi . 
\label{1} 
\end{equation} 
In this equation and in the following, the explicit space and time
dependences of the variables and parameters are implicit, unless it is
necessary or convenient for clarity. The time derivatives will be denoted
by indices $t$.
In Eq.~(\ref{1}) we are assuming dimensionless variables: the
unit of energy is $\hbar\omega/2$; the unit of length is
$\sqrt{\hbar/(m\omega)}$; and the unit of time is $1/\omega$. 
$V\equiv V(\vec{r})$ is a static trap potential, that we assume 
the harmonic oscillator with spherical symmetry and given by
$V=r^2$, in the present units.
$\lambda_2$ and  $\lambda_3 $ are the parameters of the two- and
three-body interactions, which in general can be complex quantities. 
The imaginary parts of $\lambda_2$ and $\lambda_3$ describe, respectively, 
the effects of inelastic two and three-body collisions on the dynamics of
BEC. In the present paper, we are not considering dissipative terms, and
such cubic and quintic parameters are real.
In order to compare with the formalism given in Ref.~\cite{jpb00},
$\lambda_2$ is proportional to the two-body scattering
length $a_{sc}$ and given by $\lambda_2\equiv 8\pi a_{sc}$. 

The chemical potential $\mu$ is given by the eigenvalue solutions of
Eq.~(\ref{1}), with $\psi(r,t) = e^{-{\rm i} (\mu t/2)}\varphi(r)$:
\begin{equation} 
\mu \varphi = -\Delta \varphi + r^2 \varphi + \lambda_2 |\varphi|^2
\varphi +\lambda_3 |\varphi|^4 \varphi \label{mu0}.
\end{equation} 

The Lagrangian density corresponding to Eq.~(\ref{1}) is given by
\begin{eqnarray}
{\cal L} &=& 2\;{\rm Im}(\psi^*_t \psi) - |\nabla \psi|^2 -
r^2 |\psi|^2 - 
\frac{\lambda_2}{2}|\psi|^4 - \frac{\lambda_3}{3}|\psi|^6.
\label{2}
\end{eqnarray}

\subsection{Gaussian variational approach}

To analyze the dynamics of the BEC under two and three-body interactions, 
it is convenient to follow the variational approach developed in 
Refs.~\cite{Anderson,Gaididei2}. This approach was successfully employed,
recently, in \cite{Garcia1}, for BEC with two-body interaction.
We choose the simple Gaussian ansatz 
\begin{equation}
\psi(r,t) = A(t)\exp\left( -\frac{r^2}{2a^2 (t)} + {\rm i}\frac{b(t)
r^2}{2} + {\rm i} \phi(t)\right),
\label{anz}
\end{equation}
where $A(t)$ is the amplitude, $a(t)$ is the width, 
$\phi(t)$ is the linear phase of the condensate, and
$b(t)$ is the ``chirp" parameter previously discussed.

Without dissipative terms, the normalization of the wave-function 
is conserved and given by the number of particles $N$.  
The mean-square-radius and the normalization, for a system with $D$
dimensions, are 
\begin{equation}
\langle r^2\rangle = \frac{D}{2}a^2 \;\;\;{\rm and}\;\;\;
N = A^2(\sqrt\pi a)^D = {\rm constant}.
\label{RandN}\end{equation}

For a system with $D$ dimensions and radial symmetry, the
averaged Lagrangian expression is given by
\begin{eqnarray}
-{L} = (\sqrt \pi a)^{D}A^2\left[ 
2\phi_t + \frac{D}{2}a^2 \left(b_t+\frac{1}{a^4}+b^2+1\right) +
\right.\nonumber\\
\left.+ \frac{\lambda_2A^2}{2\sqrt{2}^D} +
\frac{\lambda_3A^4}{3\sqrt{3}^D}\right], 
\label{L}
\end{eqnarray}
where 
\begin{equation}
\delta\int {L}dt = 0, \;\;\;\;
{L} = \int {\cal L}(r,t)d\vec{r}.
\label{av-L}
\end{equation}
Analyzing the corresponding Euler-Lagrange equations, 
\begin{equation}\label{EL}
\frac{\partial{L}}{\partial\eta_i} -
\frac{d}{dt}\frac{\partial{L}}{\partial\dot{\eta}_i} = 0,
\end{equation}
where $\eta_i$ refer to the variational parameters ($A(t)$, $a(t)$,
$b(t)$ and $\phi(t)$), we obtain 
\begin{equation}
a_t = a b ; \label{at}
\end{equation}
and
\begin{equation}
b_t = \frac{1}{a^4} - b^2 - 1 + \frac{\lambda_2 N a^{-(D+2)}}
{2(2\pi)^{(D/2)} } + \frac{2\lambda_3 N^2 a^{-(2D+2)}}
{3\left(\pi\sqrt{3}\right)^D }. 
\label{bt}
\end{equation}
Equation (\ref{at}) expresses the chirp parameter $b$ via initial
focusing (defocusing), $(a_t)$, of the wave-function. 
Combining the Eqs. (\ref{at}) and (\ref{bt}), we have
\begin{equation}
a_{tt} = \frac 1{a^3} - a + \frac{P}{a^{(D+1)}} + \frac{Q}{a^{(2D+1)}},
\label{att} 
\end{equation}
\begin{eqnarray}
P &\equiv& \frac{\lambda_2 N}{2\sqrt{(2\pi)^D}}\nonumber ,\\
Q &\equiv& \frac{2\lambda_3 N^2}{3\left(\pi\sqrt{3}\right)^D} =
\frac{(2)^{(D+3)}}{(3)^{(D/2+1)}}\frac{\lambda_3}{\lambda_2^2}P^2.
\label{PQ}
\end{eqnarray}

Comparing the result of $a_{tt}$ with $D=2$, from Eq. (\ref{att}), with
Ref.~\cite{Garcia1}, we observe that the resulting equation 
obtained by the variational approach coincides with the one obtained by
the moments method \cite{Vlasov}. The equation is also close to
the one obtained by the modulation method for the $2D$ Townes
soliton~\cite{Fibich}.  
In case of $D=3$, expressing the above definitions by the parameters 
given in Ref.~\cite{Gammal}, where $\lambda_2 =
8\pi a_{sc}$ in our dimensionless units ($a_{sc}$ is the two-body
scattering length), and $N = {n}/{\left(2\sqrt2 |a_{sc}| \right)}$:
\begin{equation}
|P| = \frac{n}{2\sqrt\pi}\;\;\;{\rm and}\;\;\; \ Q =
\frac{8n^2}{9\pi\sqrt {3}}g_3 . \label{PandQ} 
\end{equation} 
The anharmonic potential is derived using the mechanical
analogy. By using explicitly the dimensions in (\ref{RandN}),
\begin{equation}
R\equiv\sqrt{\langle r^2 \rangle} = \sqrt{\frac{D}{2}} a
\left(\frac{\hbar}{m\omega}\right)^{1/2} ,
\label{R2}\end{equation}
\begin{equation}
m \frac{d^2}{d t^2} R = 
\sqrt{\frac{D}{2}} a_{tt}
\left[m\omega^2\left(\frac{\hbar}{m\omega}\right)^{1/2}\right],
\label{ma}\end{equation}
a result that can be identified with 
\begin{equation}
-\frac{\partial U_P(R)}{\partial R} 
= -\left(\frac{\hbar\omega}{2}\right)
\sqrt{\frac{2}{D}} \left(\frac{m\omega}{\hbar}\right)^{1/2}
\frac{\partial U}{\partial a} ,
\label{F}\end{equation}
where 
\begin{equation}
U_P(R)\equiv U(a) \frac{\hbar\omega}{2}.
\label{UR}\end{equation}
The total energy for the BEC system with $N$ particles is
given by
\begin{equation}
E_T = N U_P(R) = N U(a) \frac{\hbar\omega}{2} \equiv E(a)
\frac{\hbar\omega}{2}.
\label{ET}\end{equation}
So, for the dimensionless quantities we obtain
\begin{equation}
\frac{\partial U}{\partial a} = -(D) a_{tt}.
\label{duda}\end{equation}
With (\ref{att}), we have the anharmonic
potential, in dimensionless units:
\begin{equation}
U(a) = \frac{D}{2}\left(a^2 + \frac1{a^2}\right) + \frac{P}{a^D}
+ \frac{Q}{2\;a^{(2D)}} .
\label{U} \end{equation}
Correspondingly, from Eq.~(\ref{mu0}),
the chemical potential is given by
\begin{equation}
\mu(a) = \frac{D}{2}\left(a^2 + \frac1{a^2}\right) + \frac{2P}{a^D}
+ \frac{3Q}{2\;a^{(2D)}} .
\label{mu} \end{equation}
 
The asymptotic limit $a\to\infty$, for $U(a)$, is the same of the
oscillator [$U(\infty)\to\infty$], and is not determined by the 
parameters $P$ and $Q$. 
So, the only limiting condition that is strongly affected by these
parameters is the limit $a=0$. When $Q\ne 0$, the three-body term
is dominant in this limit:
$ U(0) \approx {Q}/{(2 a^{2D})}.$
Now, two cases have to be considered: 
$Q<0$ give us $U(0)\to -\infty$; and
$Q>0$ will give us $U(0)\to +\infty$. 
As the other end is fixed by the oscillator condition
[$U(\infty)\to +\infty$], in the first case ($Q<0$) we cannot obtain more
than one maximum and one minimum for finite $a$, that will depend on the
sign and relative value of the parameter $P$.
The case $Q>0$ can be very interesting, if $P<0$: in both ends $U(a)$
goes to $+\infty$, such that in between we can have two minima and
one maximum for the total energy. The two minima will represent two
possible phases, and a rich dynamics can be described by the NLSE.
From this analysis, we should note that phase transition is possible
only in case $Q>0$ with $P<0$. 
When $Q\le 0$, there is only a single phase. 

\subsection{Critical parameters for stability}

The relevant extremes of $U(a)$ are given by the real and 
positive roots of 
\begin{equation}
a_s^{2} - \frac{1}{a_s^{2}} - \frac{P}{a_s^{D}} - \frac{Q}{a_s^{2D}} 
= 0\;.
\label{Droots}\end{equation}
In such extreme positions, we obtain
\begin{equation}
U(a_s) = {D} a_s^{2} + \frac{P}{2 a_s^D}\left(2 - {D}\right)
+ \frac{Q}{2\;a_s^{2D}}\left(1 - {D}\right) \; , 
\label{U2} \end{equation}
\begin{equation}
\mu(a_s) = {D} a_s^{2} + \frac{P}{2 a_s^D}\left(4 - {D}\right)
+ \frac{Q}{2\;a_s^{2D}}\left(3 - {D}\right) .
\label{mu2} \end{equation}

In Table I we resume the particular cases, from the above three
equations, where exact analytical solutions can be found for the
extremes of $U$ and the corresponding critical parameters for stability.
We note that, once a solution for a system with $D$ dimensions and
$P=0$ is known, the same solution can be used for the roots of a 
system with $2D$ dimensions and $Q=0$, by just exchanging $P$ and $Q$.
When $D=2$, the result $P_c=-1$ for the particular case when the quintic
term is zero ($Q=0$) is well known, and corresponds to $\nu=2$ in
Ref.~\cite{Gammal2}. The solutions will be unstable for $P=-|P|$ with
$|P|>1$. 

Actually, one of the interesting cases occurs when the cubic term is zero
($P=0$), for $D=1$ and $D=3$, in view of the possibility of altering the
two-body scattering length by an external magnetic field~\cite{avaria}. 
If this condition is realized ($P=0$), the effect of the three-body
non-linear term in the mean-field approximation of the condensate will be enhanced.
\begin{table}
\caption[dummy0]{
For lower dimensions, we show the results of
particular cases where exact analytical solutions can be found.
$a_s$ are the real and positive roots of the mean-square-radius
that extremize the total energy [$E_T=NU (\hbar\omega/2)]$;
$P_c$ and $Q_c$ are the critical parameters for 
stability of the NLSE in just one local minimum. 
For $D=3$, when $P$ and $Q$ are non-zero, we show just one critical 
set of parameters for the existence of two phases.
$U_c$ and $\mu_c$ are the corresponding critical values of
$U(a)$ and the chemical potential $\mu(a)$. 
}
\begin{center}
\begin{tabular}{ccccccccc}
$D$ & $(P,Q)$ & $(P_c,Q_c)$ & $a_s^4$ & $a_c^4$ & $U_c$ &
$\mu_c$ \\
\hline \hline\\
1&(0,$Q$)&$(0,-1)$&$1+Q$&0&0&${\displaystyle \frac{-1}{a_c^{2}}}$\\
2&($P$,0)&$(-1,0)$&$1+P$&0&0&$
{\displaystyle \frac{-1}{a_c^2}}$\\
2&(0,$Q$)&${\displaystyle \left(0,\frac{2\sqrt 3}{9}\right)}$& 
Eq.(\ref{Droots})
&${\displaystyle \frac 43}$
&${\displaystyle \frac{5\sqrt
3}{4}}$&${\displaystyle \frac{17\sqrt 3}{12}}$\\
3&(0,$Q$)&${\displaystyle
\left(0,\frac{-1}{4}\right)}$&${\displaystyle\frac 12\pm\sqrt{\frac 
14+Q}}$&${\displaystyle \frac 12}$
&$2\sqrt 2$&${\displaystyle\frac{3\sqrt 2}{2}}$\\
3&${\displaystyle\left(P,P^8\right)}$&${\displaystyle
\left(-\left(\frac{1}{8}\right)^{\frac{1}{4}},\frac{1}{64}\right)}$
& Eq.(\ref{Droots})
&${\displaystyle \frac 18}$
&${\displaystyle \frac{3\sqrt 2}{2}}$
&${\displaystyle\frac{-\sqrt 2}{4}}$\\
4&($P$,0)&${\displaystyle \left(\frac{2\sqrt 3}{9},0\right)}$& 
Eq.(\ref{Droots})
&${\displaystyle \frac 43}$
&${\displaystyle \frac{5\sqrt 3}{2}}$&
${\displaystyle \frac{8\sqrt 3}{3}}$\\
6&($P$,0)&${\displaystyle
\left(\frac{-1}{4},0\right)}$&${\displaystyle\frac 12\pm\sqrt{\frac 
14+P}}$&${\displaystyle \frac 12}$
&$4\sqrt
2$&${\displaystyle\frac{7\sqrt 2}{2}}$\\ \end{tabular}
\end{center}
\end{table}
In case such three-body (quintic) term is negative, as shown in Table I,
the present variational ansatz gives analytical estimates for the
critical parameters, and for physical quantities as the
mean-square-radius, energy and chemical potentials. In Section IV we
study in more detail this case.

Other particular cases occur when phase transitions are possible. As
explained before, these situations can only happen if both parameters
$P$ and $Q$ are non-zero and $Q>0$ with $P<0$. The absolute collapse is
not possible in the condensate; but, one can obtain transitions between
two-phases. This implies that, for a system with two-body attractive
interaction and three-body repulsive one, as we increase the number of
atoms in the condensate ($|P|$ increases), we can reach a critical limit
where only one phase (stable) remains. In other words, for a particular
value of $P=P_c$, the position $a$ of one of the minima of $U$ coincides 
with the maximum, and we have an inflection point of $U$.

As an example, in Table I we show one particular case where we
have
phase transition, and also analytical solution for one of the roots
of Eq.~(\ref{Droots}), when $D=3$. 
We can observe that this condition is realized if
$0 > P\ge P_c = - (1/8)^{(1/4)}$. 
When $Q=P^8$, the critical limit occurs for $P_c=-(1/8)^{(1/4)}$
and we have three-roots for $a_s$, given by 
$a_{s1}=a_{s2}=|P_c|=0.5946$ and $a_{s3}=$0.4696.

 \begin{figure}
 \setlength{\epsfxsize}{1.0\hsize}
\centerline{\epsfbox{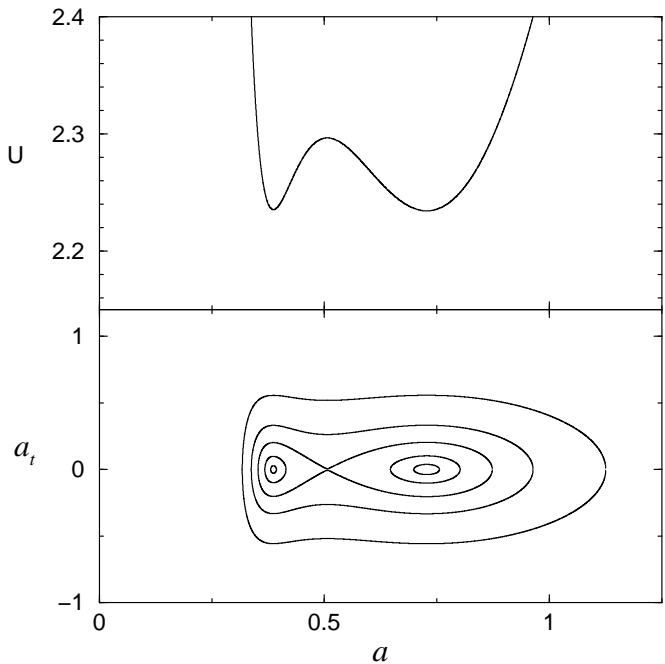}}
 \caption[dummy0]{
For a stable BEC with $\lambda_2<0$, $\lambda_3>0$ and $D=3$, 
it is shown a variational plot of the potential energy $U(a)$ (upper
frame), in units of $(\hbar\omega)/2$, 
as a function of the width $a$; and the corresponding phase plane
defined by ($a_t, a$). The parameters are dimensionless, with $P=-$0.5495
and $Q=$0.0099. $a$ is related to the mean-square-radius as
$\langle r^2\rangle = 3a^2/2$. 
   } \end{figure}

A typical plot of the potential $U(a)$ as  a function of the width $a$ and
the corresponding phase portrait defined in the plane ($a_t, a$) for
negative $\lambda_2$ and positive $\lambda_3$ is given in Fig. 1 for the
case $D=3$ of a stable BEC.
The parameters for the attractive cubic and repulsive quintic terms used
in the figure are such that $n=$1.948 and $g_3=$0.016, corresponding to 
$P=-$0.5495 and $Q=$0.0099. The number of atoms corresponds to the
situation in which we have two minima with the same value for $U(a)$. 
From the variational expressions given in this section
and our mechanical analogy, one should observe that, in  the
pictorial example of Fig. 1, if we have a solution located at the
right minimum (for example) 
it cannot migrate to the left minimum, unless the chirp parameter
corresponds to an energy greater than the difference between the
right minimum and the maximum that is in between. 

\section{Analysis of the oscillations of the BEC in three dimensions} 

The different dynamical regimes in the condensate oscillations can be
described by the cross sections of the curve of $U(a)$ with the levels of
the effective total energy. Following our mechanical analogy, we have

\begin{equation}
H(P_R,R) = \frac{P_R^2}{2m} + U_P(R)\;,
\label{H}\end{equation}
where $P_R$ is the momentum conjugate to $R$, given in (\ref{R2}).
In our dimensionless variables, the corresponding initial Hamiltonian
is given by
\begin{equation}
H_0 = \left.\frac{3}{2}(a_t)^2\right|_0 + U(a_0) =
\left(\frac{3}{2}\right)a_0^2 b_0^2 + U(a_0),
\label{H0}\end{equation}
where Eq.~(\ref{at}) was used.
Thus, by varying the initial conditions $a_0, b_0$ we alter $H_0$.

This case was recently analyzed using the moments methods in
Refs.~\cite{pita,Wadati,Berge}, under some specific assumptions.
The assumption $\left. a_t\right|_0 = 0$ was considered in
Refs.~\cite{pita,Berge}; and $\left. a_t\right|_0 \neq 0$ in 
Ref.~\cite{Wadati}. 
In Ref.~\cite{Berge}, a generalization of the Weinstein criterion 
for the collapse~\cite{weins} was obtained, in case of three 
dimensions, and when there is only cubic term in the NLSE. 

It will be interesting to derive criteria for stability
more generally applied, using a variational approach and by 
comparison with results obtained with numerical simulations.
For this purpose, we should first obtain expressions for the
mean-square-root that extremizes the total energy.
In the variational approach, for $D$ dimensions, the 
equations for the total energy (\ref{U}) and chemical potential
are given by (\ref{U}) and (\ref{mu}). So, in case of $D=3$, 
the equations corresponding to (\ref{Droots}), (\ref{U2}) and
(\ref{mu2}) are
\begin{equation}
a_s^{8} - a_s^{4} - P a_s^{3} - Q = 0
\;,
\label{3Deqr}\end{equation}
\begin{equation}
U(a_s) = {2} a_s^{2} + \frac{1}{a_s^2} + \frac{P}{2 a_s^3} 
= \frac{5a_s^{2}}{2} + \frac{1}{2 a_s^2} - \frac{Q}{2 a_s^6}\;, 
\label{U3D} \end{equation}
\begin{equation}
\mu(a_s) = {3} a_s^{2} + \frac{P}{2 a_s^3}
= \frac{7 a_s^{2}}{2} - \frac{1}{2 a_s^2} - \frac{Q}{2a_s^{6}}\;.
\label{mu3D} \end{equation}

Here we should note that an exact expression for the Hamiltonian,
corresponding to the VA given by Eqs.~(\ref{U}), can be
derived for the general case. This will be useful for the exact 
numerical calculations that we are going to perform in 3D. 
We consider the following scaling in the wave-function 
$\psi$ of the system~\cite{Kivshar}:
\begin{equation}
\psi = \alpha^{-3/2}\chi(\vec{\xi}),\;\;\;{\rm with}\;\;\;
\vec{\xi} \equiv \frac{\vec{r}}{\alpha}. 
\label{xi}\end{equation}
Then, for the total Hamiltonian, we obtain
\begin{equation}
H(\alpha) = \frac{X_s}{\alpha^2} + \alpha^2 \langle\xi^2\rangle +
\frac{\lambda_2}{2}\frac{Y_s}{\alpha^3} + 
\frac{\lambda_3}{3}\frac{Z_s}{\alpha^{6}}, \label{Ha}
\end{equation}
where 
\begin{equation}
X_s \equiv -\int\chi^\dagger \Delta \chi d^3\xi, \;\;\; 
\langle\xi^2\rangle = \int\xi^2 |\chi|^2 d^3\xi, 
\end{equation} 
\begin{equation}
Y_s \equiv \int |\chi|^4 d^3\xi ,\;\;\; 
Z_s \equiv \int |\chi|^6 d^3\xi .
\end{equation}
By taking $\delta H(\alpha) /\delta \alpha |_{\alpha=1}=0$, 
we obtain the characteristic equation
\begin{equation}
{X_s} - \langle r^2\rangle +
\frac{3}{4}\lambda_2{Y_s} + 
{\lambda_3}{Z_s} = 0\; .
\label{exroots}\end{equation} 
With the above equation, for the energy we have
\begin{equation}
H = \frac{2 X_s}{3} + \frac{4}{3}\langle r^2\rangle - 
\frac{\lambda_2}{4}{Y_s}.
\label{Has}
\end{equation}
Correspondingly, for the chemical potential,
we obtain 
\begin{equation}
\mu = 2\langle r^2\rangle
+\frac{\lambda_2}{4}{Y_s}.
\label{muas}
\end{equation}
Interesting to observe that when the two-body interaction is zero,
the exact result for the chemical potential is twice the mean 
square radius, as shown above [in agreement with the variational
result of Eq.(\ref{mu3D}), where $a_s^2=2/3\langle r^2\rangle$].
 As specific cases will be considered in
the next subsections, we will discuss this point in more detail.
For now, we can use the specific example of $Q=0$, represented
in Fig. 2, in order to illustrate the validity of a general criterion 
(also valid for $Q\ne 0$) that we consider. In this case, we have
just one minimum and a maximum of $U(a)$.

 \begin{figure}
 \setlength{\epsfxsize}{1.0\hsize}
\centerline{\epsfbox{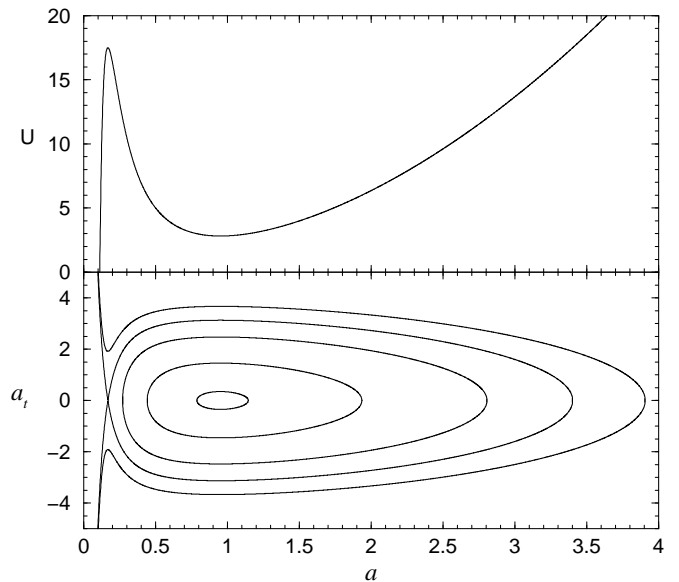}}
 \caption[dummy0]{
For $\lambda_2<0$ (with $n=0.6$, or $P=-$0.1692) and $\lambda_3=0$
($Q=0$), as in Fig. 1, we show a variational plot
of $U(a)$ (upper frame), as a function of the width $a$; 
and the corresponding phase plane defined by ($a_t, a$). 
 } \end{figure}

A given chirp $b(t)$ in the wave-function corresponds to the square
root of the kinetic energy, given by the Hamiltonian in
Eq.~(\ref{H0}). With a small initial chirp, we should have
oscillations around the minimum of $U(a)$. The square of the chirp 
must be larger than the difference between the minimum and the maximum of
the effective potential energy $U(a)$ for the system to become unstable.
So, from Eq.~(\ref{H0}) and given that $H_0$ is conserved,
we can derive a minimum criterion for stability. The given kinetic energy
(related to the chirp) must be smaller than the corresponding variation
of the potential energy, in order to maintain stable oscillations
around the minimum (where $a=a_{s1}$). The initial chirp $b_0$ must
satisfies the condition
\begin{eqnarray}
b_0^2 a_{s1}^2 &\le& -\frac{2}{3}\Delta U(a),
\nonumber \\
b_0^2 &\le& \frac{2}{3}\frac{U(a_{s2})-U(a_{s1})}
{a_{s1}^2} \; ,
\label{DH0}\end{eqnarray}
where $a_{s2}$ is the value of $a$ corresponding to the maximum of
$U(a)$, and $a_{s1}$ corresponds to the minimum of $U(a)$.
The maximum initial chirp is given by
\begin{eqnarray}
b_{0,m} &=& \sqrt{\frac{2}{3}}\frac{\sqrt{U(a_{s2})-U(a_{s1})}}
{a_{s1}} \; .
\label{b0c}\end{eqnarray}
From the above equation, we should also note that the critical
values for the number of atoms, related to the parameters $P$ and
$Q$, can be obtained by the condition that $U(a_{s2})\to U(a_{s1})$.
We should also observe that the chirp parameter is introduced
in the exact wave function by an exponential factor, 
as in the VA [see (\ref{anz})].

The frequency of the linear collective oscillations, $\omega_{L}$, is 
also another relevant quantity to analyze the stability of the
condensate. When the system 
becomes unstable, it reaches the value zero. 
From the radial variation of the force around a minimum
$a_{s1}$, obtained from Eqs.~(\ref{3Deqr}) and (\ref{U3D}), 
$\omega_{L}$, in units of the trap frequency $\omega$,
is given by
\begin{eqnarray}
{\omega_{L}} = \sqrt{\frac{1}{3}\left.\frac{d^2 U}{d
a^2}\right|_{a=a_{s1}}}
&=& \sqrt{5-\frac{1}{a_{s1}^4}+\frac{3Q}{a_{s1}^8}} \nonumber \\
&=& \sqrt{8-\frac{4}{a_{s1}^4}-\frac{3P}{a_{s1}^5}} ,
\label{wcol}\end{eqnarray}
One of the requirements for instability of the BEC is that
$\omega_{L}$ is zero, when the minimum and maximum of $U$ disappear.
So, it is convenient to analyze the critical points when the system
becomes unstable; at such points, the system collapses (in case we have
only one minimum) or one of the phases (characterized by a minimum)
disappears. From Eq.~(\ref{U3D}), 
\begin{eqnarray}
\frac{1}{3}\left.\frac{d^2 U}{d a^2}\right|_{a=a_{s}}
&=&\frac{5a_{s}^8 - a_{s}^4 + 3 Q}{a_s^8}\nonumber\\ 
&=&\frac{8a_{s}^5 - 4a_{s} - 3 P}{a_s^5} 
.\label{Qc}\end{eqnarray} 
We can only have two minima (given by $a_{s1}$ and $a_{s3}$) and one
maximum (at $a_{s2}$) for negative $P$ and positive $Q$. In this
case, two critical limits are possible, corresponding
to the situations in which the position of one of the minima, $a_{s1}$ or
$a_{s3}$, is equal to the position of the maximum (one of the phases
disappears).
Note that in the critical limit we have $a_{min} = a_{max} = a_c$.
In the next two subsections, we will consider the particular cases
in which we have only one nonlinear term present in the NLSE: the
cubic term or the quintic term. The discussion of the more general cases,
where
both terms are present in the NLSE, we leave for the section IV.

\subsection{Case of $Q=0$, with attractive two-body term}

The roots for the maximum ($a_{s2}$) and the minimum ($a_{s1}$) of $U$ 
can be obtained from (\ref{3Deqr}):
\begin{equation}
a_s^5-a_s+|P| = 0\; .
\label{Q0}\end{equation}
Combining this equation with Eq.~(\ref{Qc}), we have
\begin{eqnarray}
\frac{1}{3}\left.\frac{d^2 U}{d a^2}\right|_{a=a_{s}}
&=& 5 - \frac{1}{a_s^4} 
.\label{Pc}\end{eqnarray} 
In the critical limit~\cite{stoof2,fetter},
\begin{eqnarray}
a_c &=& \left(\frac{1}{5}\right)^{\frac{1}{4}}, \;\;\;
P_c = -\frac{4}{5}\left(\frac{1}{5}\right)^{\frac{1}{4}}
= - 0.53499 \label{acQ0} \\
{\rm and}\;\;\;\;  
n_c &=& 2\sqrt{\pi}|P_c| = 1.8965 ,\nonumber 
\end{eqnarray}
where Eqs.~(\ref{Q0}) and (\ref{PandQ}) were used.
With the above root of Eq.~(\ref{Q0}), the other roots can
also be easily found numerically for any value of $P$. 
Given $P$, once we obtain the positions of the minimum 
($a_{s1}$) and maximum ($a_{s2}$) of $U$, we can calculate
the value of the maximum initial chirp for the collapse,
using Eq.~(\ref{b0c}):
\begin{equation}
b_{0,m} = 
\sqrt{\frac{a_{s1}^2 -
a_{s2}^2}{3a_{s1}^4a_{s2}^2}(1-5a_{s1}^2a_{s2}^2)}
.\label{bcQ0}\end{equation}
Our results, when we have only the cubic term in the
nonlinear effective potential ($Q=0$), is shown in Figs. 2 
and 3. In Fig. 2, for $P=-0.1693$ (corresponding to $n=0.6$),
we represent in the lower frame the phase space given
by $a_t$ versus $a$. In the upper frame, we have the total energy, in 
units of $N(3/2)\hbar\omega$, as a function of $a$. 

 \begin{figure}
 \setlength{\epsfxsize}{1.0\hsize}
\centerline{\epsfbox{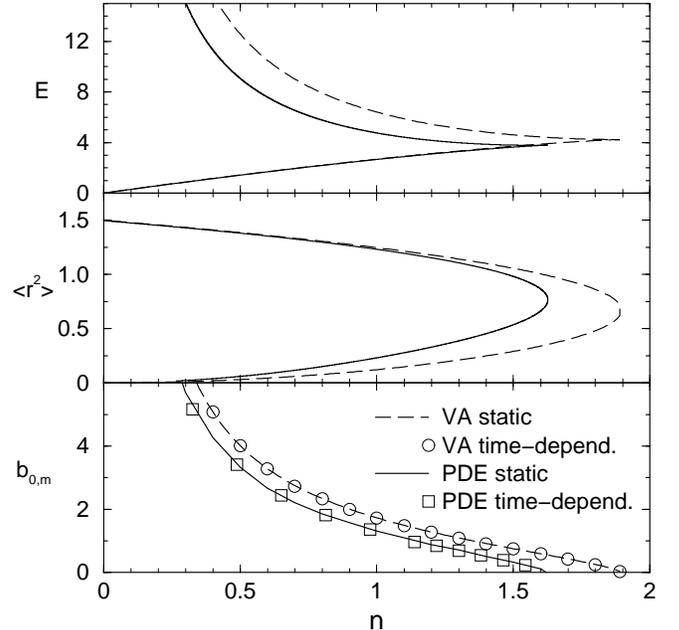}}
 \caption[dummy0]{
Results for $Q=0$, as functions of $n$, the reduced number of atoms
defined in the text. In the upper frame we have $E = n U$, the total 
energy in units of $(N/n)(\hbar\omega/2)$; in the middle frame, the
mean-square-radius $\langle r^2 \rangle$, in units of $\hbar/(m\omega)$;
and, in the lower frame, the maximum initial chirp. 
In the critical limit, $n_c=$ 1.8965 and $P_c=-$ 0.53499, the total energy
is given by $E=nU = n_c\sqrt 5 = 4.2407$, corresponding to  $E_T=\sqrt{5}
N_c(\hbar\omega/2)$.
}
\end{figure}
  
In Fig. 3, we show in the lower frame our results for the maximum
initial chirp, $b_{0,m}$, as a function of $n$, for stable solutions.
$n$ is the reduced number of atoms, given by Eq.~(\ref{PandQ}).
In this figure, we also represent the total energy (upper frame)
and the average of the square radius, $\langle r^2\rangle$, as a
function of $n$. 

The static variational expression for the maximum initial chirp, was
derived from Eq.~(\ref{at}) and energy conservation requirements, given
by Eq.~(\ref{DH0}). The circles correspond to results obtained by the
time-dependent VA, by increasing the initial conditions for the chirp
till the system collapses. 
We also obtain the results for the maximum initial chirp using the 
same Eq.~(\ref{DH0}), but considering exact PDE results for the
observables presented in it.
The results, for the static case, are represented by a solid line;  
and, for the time-dependent case, by squares.
We have observed small differences in the absolute values of $b_{0,m}$
when considering the trial $b_{0}$ as positive or negative. 
This fact can be 
qualitatively understood by a small dephasing of the wave that 
occurs near the wall of the oscillator, when it is bouncing back.

For the mean-square-radius, we observe that the VA results
approaches the exact PDE results also in the unstable branch in 
the limit $n\to 0$. These results are consistent with the ones
obtained in Ref.~\cite{Kivshar}, where it is shown that, for the 
trapped NLSE with cubic term, the unstable branch of the chemical
potential diverges in the limit $n\to 0$, with $\langle r^2\rangle$
collapsing to zero (this limit is the non-trapped solution of 
the NLSE). 
 
\subsection{Case of $P=0$ with attractive three-body term}

As in the previous subsection, for a stable condensate the maximum
initial chirp is given by Eq.~(\ref{b0c}), where the roots 
are given by Eq.~(\ref{3Deqr}). The real and positive
roots correspond to stable solutions, such that only one
is meaningful when $Q>0$. When $Q<0$, we obtain two real and positive
roots of $a$, in case $0>Q\ge -1/4$. 
These roots are 
\begin{equation}
a_{{s1}/{s2}}\equiv
a_{+/-}=
\left[\frac{1}{2}\pm\sqrt{\frac{1}{4}+Q}\right]^\frac{1}{4},
\label{3roots}\end{equation}
where the minimum of $U$ is given by $a_+$ ($\equiv a_{s1}$) and the
maximum by $a_-$ ($\equiv a_{s2}$).
The critical limit happens for $a_c^4 = 1/2$, or $Q_c=-|Q_c|=-1/4$.
From Eqs.~(\ref{3roots}) and (\ref{Qc}), 
\begin{equation}
\frac{1}{3}\left.\frac{d^2 U}{d a^2}\right|_{a=a_{s}}= 
\pm\frac{2\sqrt{1-4|Q|}}{|Q|}
\left(1\mp\sqrt{1-4|Q|}\right),
\label{u2}\end{equation}
and the frequency of the linear
collective oscillations, $\omega_{L}$, near the minimum 
is given by
\begin{equation}
\omega_{L} = 
\sqrt{
\frac{2\sqrt{1-4|Q|}}{|Q|}
\left(1-\sqrt{1-4|Q|}\right)
}.
\label{wcolP0}\end{equation}

In Fig. 4, we present the main results for this subsection with $P=0$.
As noted in Eqs.(\ref{mu3D}) and (\ref{muas}) (respectively, the
variational and exact expressions), the chemical potential is just 
twice the mean-square-radius. In the plots, the VA results are
shown in dashed lines and the exact PDE in solid lines.
The variational critical limit, $Q=Q_c=-1/4$ ($a_c^4=1/2$), corresponds to 
$n_{3,c} = 1.2371 \;,$ where $n_{3}$ is the reduced number of atoms,
that is given by the normalization of the wave-function. In this case, 
Eq.~(\ref{PQ}) will give us the relations between $|Q|$,
$|\lambda_3|$ and $n_3$:
\begin{equation}
n_{3} = \frac{\sqrt{|\lambda_{3}|} N}{2\pi}\;\;\;{\rm and}\;\;\;
|Q|=\frac{8 n_3^2}{9\pi\sqrt 3}.\label{Qn3}  \end{equation}
Given $\lambda_3$, 
we can obtain the corresponding critical number of atoms:
\begin{equation}
N_c = \frac{3\pi^{3/2} 3^{1/4}}{2\sqrt{2}\sqrt{|\lambda_3|}}
 = \frac{7.7729}{\sqrt{|\lambda_3|}},\label{Nc}\end{equation}
a result consistent with no limit in the number of atoms if
$\lambda_3\to 0$.

 \begin{figure}
 \setlength{\epsfxsize}{1.0\hsize}
\centerline{\epsfbox{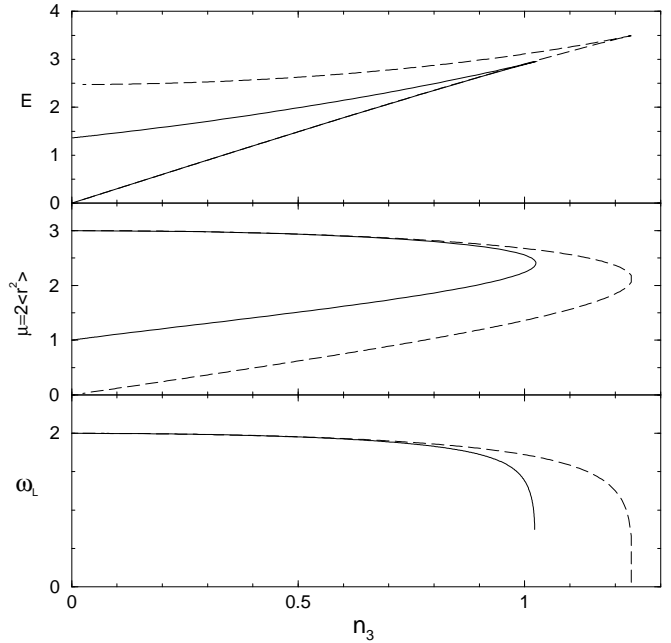}}
 \caption[dummy0]{
Results for $P=0$, as functions of the reduced number of atoms 
$n_3$. The total energy is given in the upper frame,
in units of $(N/n_3)(\hbar\omega/2)$; in the middle frame we have the 
chemical potential in units of $(\hbar\omega)/2$; and, in the lower frame, 
the frequencies of the collective breathing mode. 
In the VA, the critical limit for stability, $Q=Q_c=-1/4$, 
corresponds to $n_{3,c} = $1.2371, where $E=n_3 U_c = 2\sqrt 2 n_{3,c} =
3.499$ [corresponding to  $E_T=2\sqrt{2} N_c (\hbar\omega/2)$].
} \end{figure}

In Fig. 4, we also observe that $\partial \mu /\partial n_3< 0 $
(branch of minima)  and $\partial \mu /\partial n_3> 0 $ (branch of
maxima)
correspond, respectively, to stable and unstable solutions. 
We note that localized stable structure can only exist for positive
$\mu$. The results in this case, that the effective interaction 
contains only the trap and quintic term, are in agreement
with the VK criterion~\cite{VK}. 
The observed exact relation between the chemical potential and the
mean-square-radius ($\mu=2\langle r^2\rangle$), in the present
case that $\lambda_2=0$, was proved 
in Eq.~(\ref{muas}).

In the branch with the unstable solutions, we observe numerically
that in the limit $n_3 \rightarrow 0$ the wave-function approach the 
form $\psi \propto \exp(-r^2/2)/r$; with $\mu\to 1$ and $\langle
r^2\rangle\to$1/2. This solution is proved analytically to be the
irregular solution of the oscillator [Recall that, when $\lambda_2=0$
and $\lambda_3=0$, in Eq.(\ref{mu0}), the regular oscillator solution 
gives $\mu = 3$; as for the irregular solution in the origin we have
$\mu = 1$ in our energy unit ($\hbar\omega/2$)].
So, for the unstable solutions, with small $|Q|$, the Gaussian ansatz
(started by considering the exact regular solution of the oscillator in
the limit $n_3=0$) fails. 
By keeping (artificially) the Gaussian shape for the unstable solutions,
the radius is forced to be zero.
This explains the discrepancy between the present VA and the PDE results
in the unstable region. The more appropriate ansatz for the trapped NLSE 
(when $\lambda_2$ and/or $\lambda_3$ is nonzero) should be build 
considering the two (regular and irregular) solutions of the harmonic
oscillator. 

Even considering the discrepancy between VA and PDE results for the
unstable (maxima) solutions, there is a reasonable agreement for the
results obtained for the maximum initial chirp $b_{0,m}$, as shown in
Fig. 5. In Fig. 6, we have the variational result for the potential
energy as a function of $a$, and the corresponding phase space, as in
Figs. 1 and 2.
Static predictions on PDE are also in good agreement for the maximum
initial chirp. In PDE calculations we have verified numerically small
differences between negative and positive $b_{0,m}$, as in the case
analyzed and explained in the previous subsection (related to Fig. 3).
Using Eq.~(\ref{b0c}) with $P=0$, the maximum initial chirp
is given by
\begin{eqnarray}
b_{0,m} &=& \sqrt{
\frac{1}{3|Q|}
\left(1-2\sqrt{|Q|}\right)\left[\sqrt{1-4|Q|}-1+2\sqrt{|Q|}\right]
}\nonumber \\
&\approx& 
\sqrt{\frac{2}{3 \sqrt{|Q|}}}\left(1- \frac{3\sqrt{|Q|}}{2}\right)
\;\;\;{\rm for}\;\;\;|Q|<<1 .
\label{bcP0}\end{eqnarray}
The above result shows that, when the three-body parameter is
negative and small, the maximum initial chirp is 
$\propto |Q|^{-1/4}$, or $\propto (1/\sqrt n_3)$, as shown in 
Fig. 5.

 \begin{figure}
 \setlength{\epsfxsize}{1.0\hsize}
\centerline{\epsfbox{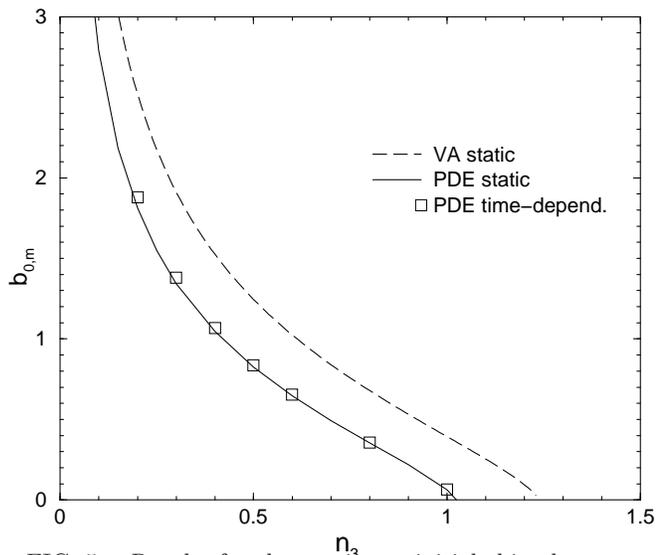}}
\vskip -0.3cm
\caption[dummy0]{
Results for the maximum initial chirp $b_{0,m}$,
considering $\lambda_3<0$ and $\lambda_2=0$, using
the VA and exact PDE results.
} \end{figure}
\vskip -0.5cm
 \begin{figure}
 \setlength{\epsfxsize}{1.0\hsize}
\centerline{\epsfbox{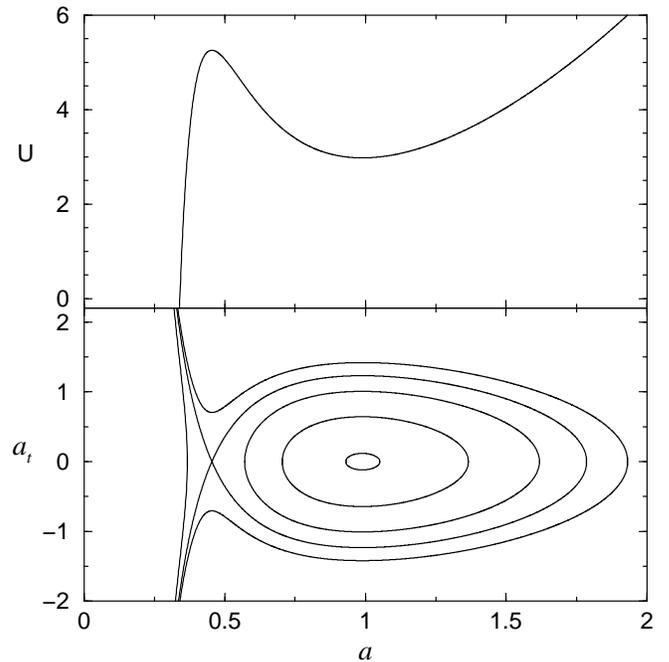}}
\vskip -0.2cm
 \caption[dummy0]{
For $\lambda_3<0$ and $\lambda_2=0$ ($Q<0$ and $P=0$), with $n_3=$0.5,
as in Figs. 1 and 2, we present results of the VA for
the total potential energy $U$ as a function of $a$, and a plot of the
phase space of $a_t$ versus $a$.
 } \end{figure}

\section{Dynamics of BEC with two- and three-body interactions}

In this section, we analyze the cases where we have non-zero 
cubic ($\lambda_2$) and quintic ($\lambda_3$) interaction terms 
in the NLSE. 

The different possibilities, relative to the signs of the 
two and three-body interactions are studied in the next three
sub-sections, where we exclude one of the cases in which
the signs are the same. When $Q$ has the same sign of $P$, the basic
physical picture is not essentially altered in comparison to the
cases already discussed, where one of these quantities is zero. 
However, it is worthwhile to examine the changes in the 
collective excitations in the case that $\lambda_2$ and 
$\lambda_3$ are positive, as such observable can be useful to
obtain informations about possible manifestation of three-body
interactions.

\subsection
{The case of repulsive cubic and quintic interactions
($\lambda_2>0$ and $\lambda_3>0$)}

From Eq.~(\ref{wcol}), one should observe that
the frequency of the collective oscillations will decreases, as we 
increase $P$, or if the position of the minimum, $a_{s1}$, decreases. 
As all quantities present in Eq.~(\ref{wcol}) are positive, this
frequency in the VA has an upper limited at the value $2\sqrt 2$, such
that a reasonable frequency of oscillations around the minimum will be
smaller than that (see also Ref.~\cite{Stringari}).

The effects of the unharmonicity of oscillations can be taken into account
by the expansion of the effective potential  near the bottom $a_c$ on
the power of deviations $y = a - a_{s1}$  .
The equation for $y(t)$ is:
\begin{equation}
y_{tt} = -B_1 y - B_2 y^2 - B_3 y^3
\end{equation}
where 
\begin{eqnarray}
B_1 = 1 + \frac{3}{a_{s1}^4} + \frac{4P}{a_{s1}^5} +
\frac{7Q}{a_{s1}^8},
\nonumber \\
B_2 = -\frac{6}{a_{s1}^5} - \frac{10P}{a_{s1}^6} -
\frac{28Q}{a_{s1}^9}
,\nonumber \\
B_3 = \frac{10}{a_{s1}^6} + \frac{20P}{a_{s1}^7} +
\frac{84 Q}{a_{s1}^{10}}
.\label{AAA}\end{eqnarray}
Then, applying the perturbation techniques to the theory of 
nonlinear oscillations~\cite{Landau}, we find the frequency of
weakly nonlinear collective excitations:
\begin{equation}
\omega_{NL} = \sqrt{B_1}\left[1 + \left( \frac{3B_3}{8 B_1} -
\frac{5B_2^2}{12 B_1^2} \right)\sigma^2\right] ,
\end{equation}
where $\sigma$ is the amplitude of the oscillations. The correction
to linear frequency is proportional to the square of the amplitude of
oscillations of the condensate. 

The above estimate of the nonlinear oscillations in BEC, when both
the two and three-body terms are positive, can be a relevant
piece of information to determine a possible manifestation of 
three-body interaction in the condensates and its magnitude.
We should observe that, recently, in Ref.~\cite{Hec}, it was
reported observation of nonlinear oscillations in BEC of a gas with
rubidium atoms.

Let us estimate the amplitude of oscillations of the width of the atomic
cloud. The points of maximum and minimum width are defined by the section
points of the line $H_0 = const$ with the potential curve $U(a)$. For
large width, $a >> 1$, we can obtain the estimate
\begin{equation}
\sigma_{\pm} = \sqrt{\frac{H_0}{3} \pm
\sqrt{\left(\frac{H_0}{3}\right)^2 -
1}},
\end{equation}
where $\sigma_{+(-)}\equiv \sigma_{max(min)}$.
It is natural to assume that the asymptotic value of the width is the
averaged value between $\sigma_{max}$ and $\sigma_{min}$.

\subsection
{Case of attractive cubic and repulsive quintic interactions
($\lambda_2<0$ and $\lambda_3>0$)}

In this case, there is no collapse, but the condensate can
have up to two distinct phases  when $|\lambda_2|$ is smaller than a
critical value. 
From the beginning, when $\lambda_3=0$, $U$ has one minimum at
$a=a_{s1}$ and a maximum at $a=a_{s2}$. By considering a
fixed positive (and small) $\lambda_3$, a second minimum of $U$
($a_{s3}$) appears, corresponding to a denser phase in the condensate. 
As we increase the value of $|P|$, we can reach a critical
value where the minimum of the normal phase disappears~\cite{jpb00}.

When $H_0/3 < \Delta U = |U(a_{s2}) - U(a_{s})|$, with 
$a_{s}=a_{s1}$ or $a_{s3}$, 
we have small amplitude oscillations near the fixed point $a_s$. 
With $H_0/3 > \Delta U$, the character of the oscillations changes
and we have large amplitude oscillations , due to the motion of the
effective particle between the wall given by the repulsive three-body
interaction and the other curve given by the quadratic potential.
With $a_s $ close to $a_{s3}$ and $H_0/3 < \Delta U_2 = 
|U(a_{s2} ) - U(a_{s3} )|$,
the condensate oscillates with the amplitude restricted from below by
$a_{s2}$. 

The position of the minimum $a_{s1}$ of $U(a)$ is defined by the
equilibrium between two and three-body interactions; the positions of
the other extrema of $U$, the fixed points $a_{s2}$ and $a_{s3}$ are
mainly defined by the contributions of the two- and three-body terms. 
These considerations allow us to obtain analytical expressions for
such fixed points, that are given by the real and positive roots of
the equation given by the first derivative of $U(a)$.

In order to analyze the frequencies of the linear oscillations
[given by Eq.~(\ref{wcol})], we return to Eq~(\ref{Qc}), where $P=-|P|$
and $Q=|Q|$. The physically relevant situations occur when $|P|$ is
a fraction of one and $Q << 1$. And of particular interest are the cases
where two phases (corresponding to two minima) are possible, at
$a=a_{s1}$ and $a=a_{s3}$. 

Let us consider, for example, the particular case examined in Table I,
that we have a one-point solution given by $a_{s1}=|P|$ and
$Q=a_{s1}^8=|P|^8$.
This example is chosen not only due to convenience but because it is not
far from a more realistic situation. Usually we scale the normalization
of the wave-function with $n$ that is directly proportional to $|P|$.
So, the three-body parameter $\lambda_3$ is related to the particular
point we are considering.
The critical limit, in this case, is given by $|P_{c1}|^4=(1/8)$
($|P_{c1}|=$0.5946) and $Q_{c1} = 1/64$, as
shown in Table I, corresponding, respectively, to $n_{c1}=2.1078$ and
$g_3=0.021$ [see Eq. (\ref{PandQ})].  At such critical limit the
frequency of the linear oscillations around the minimum $a_{s1}$ goes to
zero [In a practical situation, before the system reaches this 
critical limit in the normal phase, it tunnels to the denser phase,
when both energies are equal]. 
With $|P|$ larger than $|P_c|$, there is only one phase (minimum)
corresponding to a denser phase.
As we reduce the value of $|P|$, we reach a second critical point at
$a_{s2}=a_{s3}=|P_{c3}|$, where $a_{s3}$ is the position of the minimum
of the denser phase. A given value of $|P|$ should be in between the
limits $|P_{c3}|<|P|<|P_{c1}|$ for the existence of two phases.

So, for the normal phase, we note that the corresponding frequencies 
of the oscillations are approximately given by the harmonic oscillator,
with $\omega_{L1}\sim 2$ (twice the trap frequency).
This can be seen in the previous cases where one of the nonlinear
terms is zero and the other is small.

Consider the case exemplified in Fig. 1, where $P\approx -0.55$
($n\approx$1.95) and $Q\approx$ 0.01 ($g_3=0.016$), and the
corresponding roots for the minima are $a_{s3}\approx$0.386 and
$a_{s1}\approx$0.726. Using Eq.~(\ref{wcol}), we can observe that the
frequency of the oscillations in the denser phase, $\omega_{L2}=
4.493$, is much larger than the frequency of the normal phase,  
$\omega_{L1}= 1.339$.
If we take more realistic parameters as, for example, $g_3 = 0.001$ and
$n = 1.5$, using Eqs.~(\ref{PandQ}) and (\ref{3Deqr}), we obtain 
$a_{s1} = 0.840$ (minimum of the `gas' phase), $a_{s2} = 0.434$ 
(maximum) and $a_{s3} = 0.105$ (minimum of the `liquid' phase).
The corresponding frequencies of the linear oscillations of the 
two phases are $\omega_{L1} = 1.73$ and $\omega_{L2} = 258$. So, 
in general we observe that it should be expected 
\begin{equation}
\omega_{L2} \gg \omega_{L1}.\end{equation} 
This unequality  for the frequencies was also confirmed in
numerical PDE calculations done in Ref.~\cite{jpb00}. 

The amplitude of the nonlinear oscillations in the dense phase can be
found from the observation that $a_{s3} << 1$. Then the terms with
$a^{-6}$ and $a^{-3}$ are dominant in the effective potential energy,
and an approximate solution can be found:
\begin{equation}
\sigma_{max,min} = \left(\frac{|P|}{2|H_0|} \pm
\sqrt{\frac{P^2}{4H_0^2} - 
\frac{Q}{2|H_0|}}\right)^{1/3} .
\end{equation} 

An interesting phenomenon occurs when the initial $\left. a_t\right|_0 =
a_0b_0$ is large. Then, the character of cloud oscillations will change,
from oscillations near $a_{s1}$ and $a_{s3}$ to large amplitude
oscillations defined mainly by the quadratic potential (see Fig. 1).
Let us estimate the criterion for the bifurcation phenomenon in the
oscillations. 
Taking into account the expression for the energy $H_0$, the bifurcation
point is given by
\begin{equation}
b_{0,m} \ge \sqrt{\frac{2\Delta U}{3}}\frac{1}{a_{s1}}. 
\end{equation}
The result is shown in Fig. 7, using our Gaussian VA, which gives us
a qualitative picture of the exact results, that were already 
presented in Ref.~\cite{jpb00}.

\begin{figure}
\setlength{\epsfxsize}{1.0\hsize}
\centerline{\epsfbox{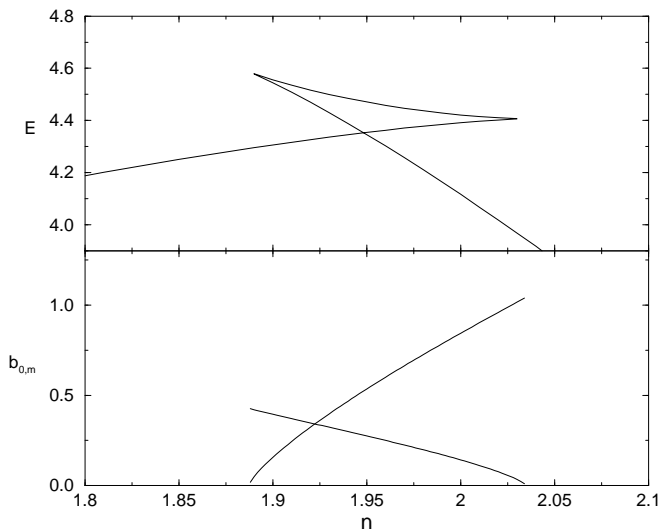}}
\caption[dummy0]{
VA results for the total energy and for the
corresponding maximum initial chirp, as functions of the reduced number of
atoms $n$, in case we have attractive two-body and a fixed
repulsive three-body interactions with $g_3=$0.016.
} \end{figure}

 \begin{figure}
 \setlength{\epsfxsize}{1.0\hsize}
\centerline{\epsfbox{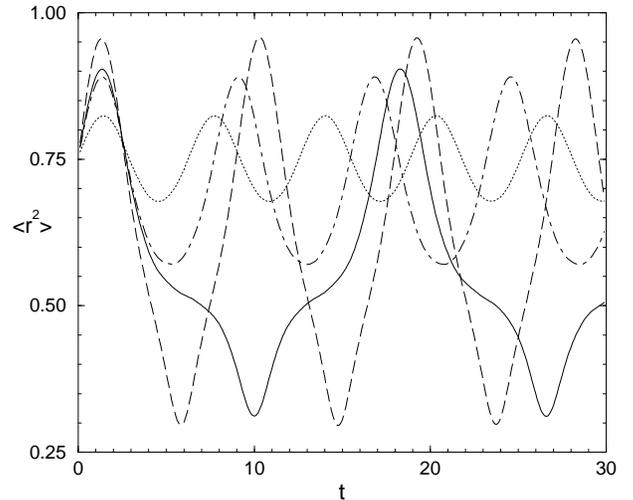}}
 \caption[dummy0]{
Time evolution of the square radius with the full PDE in
the case of phase transition for $n=$1.754, $g_3=$0.016 ($\lambda_2<0$,
$\lambda_3>0$), beginning in the gas (less dense) phase solution. The
initial conditions (chirp) considered are $b_0=$0.05 (dotted), 
$b_0=$0.10 (dot-dashed), $b_0=$0.11 (solid line), $b_0=$0.15 (long
dashed).
} 
\end{figure}
In Fig. 8 we show full numerical PDE calculation for the system
beginning in the gas phase, which corresponds to the right minimum of the
Fig. 1. Our results show that for a small chirp $(b_{0}\le 0.10)$, the
mean
square radius oscillates with small amplitude, according to its collective
frequency. Applying a stronger chirp, near $b_0\ge 0.11$, we observe
transitions back and forth between two phases, characterized by different
amplitudes of the oscillations. As $b_0$ is large enough we observe that
the oscillation patterns remains with almost fixed and large amplitude.
A similar picture happens if we begin in the denser phase, with 
different initial conditions given by the chirp. These results were 
obtained by using the exact PDE calculation, and they are in
qualitative agreement with the VA approach. The quantitative agreement
between the VA and the exact PDE deviates considerably in the phase
transition region. This should be expected, as we are far from the
harmonic oscillator behavior, in a region where an improvement in the
ansatz is necessary.

\subsection
{Case of repulsive cubic and attractive quintic interactions
($\lambda_2>0$ and $\lambda_3<0$)}

As both two and three-body nonlinear terms are 
non-zero and with opposite signs (with attractive three-body
and repulsive two-body interactions), one could expect a behavior
similar to the case that was analyzed in the previous subsection,
and also represented in Fig. 1 (where the two-body interaction is
attractive and the three-body is repulsive). But contrary to such
expectation, this case shows a different behavior for small radius,
and no phase transition is possible. We observe that in this case
the system can collapse, as the behavior for small radius is
dominated by the three-body term, that is negative. This is
represented in Fig. 9, in a variational plot of $U(a)$ versus $a$,
together with a corresponding plot for the phase space.

In Fig. 10, we present the breathing mode collective excitations for
a few values of $g_3$ calculated in the variational approximation.
The collective excitations show that even for small negative $g_3$ a
limited number of atoms is allowed in the condensate. 
Now, only a region of stationary condensate - the denser phase -
exists. Two points exist that extremize the energy: one, 
$a_{s1}$, is stable; and the other, $a_{s2}$, is unstable.
Here, the situation is similar to the cases that we have a single 
attractive nonlinear term (cubic or quintic); and the conditions for 
collapse, in terms of the initial chirp $b_0$, is such that 
it must be larger than $b_{0,m}$ given in Eq.~(\ref{b0c}).

One should note that this particular case ($\lambda_2>0$ and
$\lambda_3<0$) can be relevant in the analysis of experiments with
BEC performed with atomic systems that have repulsive two-body
interaction. No collapse is expected if a real three-body
effect is not manifested, or in case the possible three-body effect
is also repulsive.  However, an attractive three-body effect will
change this scenario, as the system must collapse for certain
critical maximum number of atoms.
In this perspective, the present analysis shows that experiments
with BEC may be useful in detecting negative three-body forces.
The corresponding maximum critical number of atoms for 
stability of the condensate can also be obtained from the present
approach, by taking the limit $b_{0,m}\to 0$ (or $a_{s2}\to a_{s1}$). 
For instance, when $g_3=-0.5$, the critical number $n$ can be obtained
from the rhs (positive $\lambda_2$) of Fig. 11.

 \begin{figure}
 \setlength{\epsfxsize}{1.0\hsize}
\centerline{\epsfbox{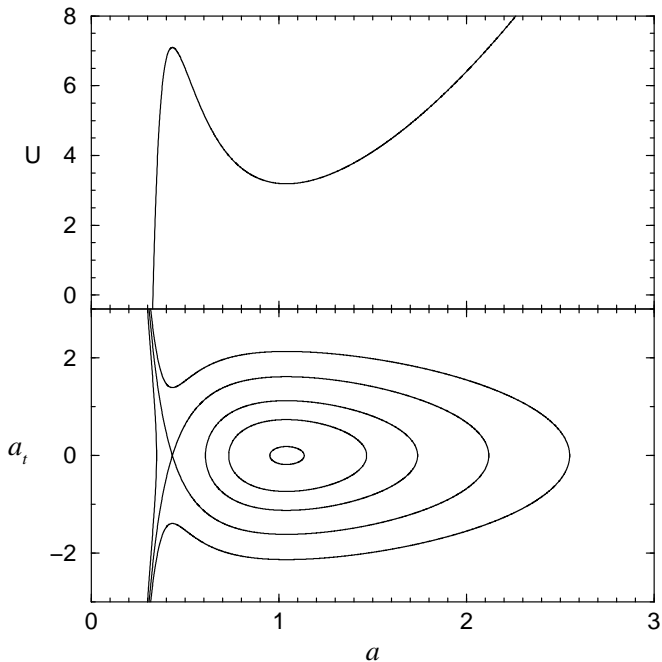}}
 \caption[dummy0]{
As in Figs. 1, 2 and 5, we have in the upper frame the potential   
energy $U(a)$; and, in the lower frame, a plot of the corresponding
phase space of $a_t$ versus $a$. We have nonzero values for both two and
three-body parameters, with $\lambda_2>0$ and $\lambda_3<0$ ($g_3=-0.5$).
}
\end{figure}

 \begin{figure}
 \setlength{\epsfxsize}{1.0\hsize}
\centerline{\epsfbox{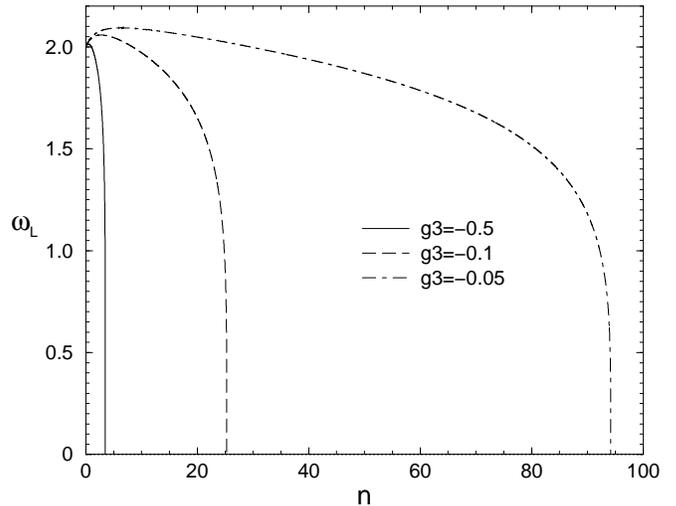}}
 \caption[dummy0]{
Collective excitations for the breathing mode of the condensate for
positive two-body interaction ($\lambda_2> 0$),
and for a set of negative three-body interaction ($\lambda_3 < 0$), as 
shown inside the figure.
}
\end{figure}

 \begin{figure}
 \setlength{\epsfxsize}{1.0\hsize}
\centerline{\epsfbox{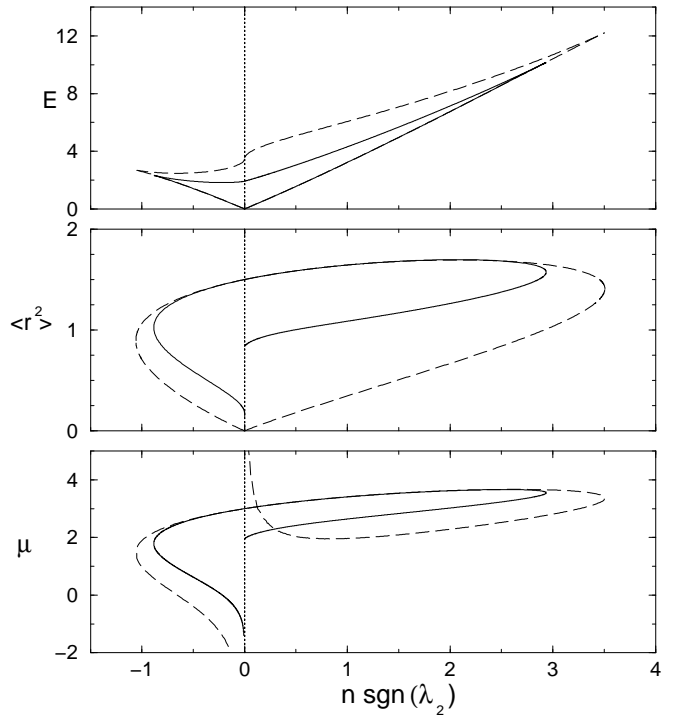}
}
 \caption[dummy0]{
Total energy $E$, mean-square-radius $\langle r^2\rangle$, and 
the chemical potential $\mu$, for negative $\lambda_3$ (with
$g_3=-$0.5). The dashed line corresponds to the variational approach
and the solid line to exact numerical calculations.
Departing from $\mu=$3, to the right or to the left, we have
the solutions corresponding to minima of the total energy, until critical
values $n_c$ (in both sides) are reached. After the critical numbers, the
curves follow lower branches, corresponding to maxima. 
} \end{figure}

We present in Fig. 11, for a fixed attractive three-body interaction,
with $g_3=-0.5$, the total energy $E$, the mean-square-radius
$\langle r^2\rangle$, and the chemical potential, as a function of the
reduced number of atoms $n$ multiplied by the sign of the two-body 
interaction. We consider both negative ($\lambda_2<0$) and positive
($\lambda_2>0$) two-body interactions, and the value of $g_3=-0.5$ was
chosen due to numerical convenience. 
The variational approach (dashed line) gives a good description of the
minima for small $|n|$, but fails particularly for the maxima solutions,
as one can see in the figure when comparing with the exact results
(solid lines). In agreement with the variational results of collective
excitations, the number of atoms allowed is limited to a critical number
$n_c\sim$ 3.5 (the corresponding exact result is $\sim $2.93).
For a more realistic value of the three-body parameter, with 
$g_3 =-$0.01, the exact critical number is $n_c \sim $4200.   

In the upper frame of the Fig. 11, it is shown a similar pattern for
the minimum and the maximum of the energy, as already
described with $\lambda_3 \geq 0$ when $\lambda_2 < 0$  and
with $\lambda_3 < 0$ when $\lambda_2 = 0$.  Such results also indicates 
that an initial wave function with enough larger chirp can make the 
condensate unstable and that the corresponding magnitude of $b_{0,m}$ can
be inferred from the absolute difference between the maximum
and minimum of the energy. This implies that, when we have positive 
two-body interaction ($\lambda_2 > 0$), the presence of a negative
three-body term ($\lambda_3<0$) can be detected in principle even if
the number of atoms has not achieved the critical number. The initial
maximum chirp plays a relevant role in this case.

\section{VK criterion} 

In the context of stability analysis of NLSE bound states, the orbital
stability, or stability with respect to the form, was first settled by the
so-called Vakhitov-Kolokolov criterion~\cite{VK}, that is detailed in a
recent review in Ref.~\cite{BergeR}. 
A solitary wave is orbitally stable if the initial orbit,
chosen near the ground-state orbit, implies that the orbit of the 
solution at any $t>0$ remains close to the ground state. 
The criterion was first demonstrated in Ref.~\cite{weins2}, 
by minimizing the deviations between the orbital states and the
ground-state, with respect to the initial position parameters.
In our notation, the VK criterion for stability can be written as
\footnote{We should note that $\mu$ corresponds to 
$-\lambda$ of Ref.~\cite{BergeR}.}
\begin{equation}
\frac{\partial N}{\partial \mu} < 0 
.\label{VK}
\end{equation}

However, the original VK criterion and the above cited 
demonstration have only considered cases without trap,
with nonlinearities expressed by $f(|\psi|^2)$. 
More recently, the general conditions that guarantees orbital
stability of stationary trapped condensates, described by the NLSE
with cubic term, was formally demonstrated by Berg\'e et
al.~\cite{Kivshar}, in agreement with results analyzed in
Ref.~\cite{stoof} for the case that the cubic term is negative.

The validity of the VK criterion appears to be solidly consolidated
by the above described formal demonstrations and by the specific cases
that have so far been considered. These results appear to support
an extended range of applicability of the VK criterion to the NLSE.
In the several cases that we have considered in the present paper
(by exact numerical procedure and also by the Gaussian variational
approach), the applicability of the VK criterion is out of the question in
the mentioned cases that have been examined by other authors. 
We have also confirmed the applicability of the criterion
in the case that we have only the trap and the quintic term in the
effective interaction, as shown in Fig.~4.

The validity of the VK criterion cannot be extended to trapped NLSE
with two and three-body terms, when the two body term is negative,
as one can observe from the results obtained in Ref.\cite{jpb00}.
This is an unexpected result, considering that the criterion is
applicable for non-trapped systems
(with cubic and/or quintic terms) and also for the trapped cases when
we have only the cubic or the quintic terms (and also in the cases
that both terms have the same sign). 
The VK criterion is also not applicable when $\lambda_2>0$, 
as observed in Fig. 11, in the right side of the plot. 
The criterion fails particularly in the region near the oscillator
solution (where $\mu=$3), where we can observe that
$\partial\mu/\partial n >0$ in a stable branch
(upper) and also in the unstable branch (lower). (We recognize the stable
branch as the one that corresponds to minima of the energy.)
Therefore, in the present work, we confirm the conclusion about the
limitation of the VK criterion: The nonvalidity of the VK criterion
is verified numerically when we have a positive harmonic trap, with
two and three-body nonlinear terms with opposite signs. 

\section{Conclusions}
The stability of a trapped condensate, with two- and
three-body nonlinear terms, was studied in the present work,
considering several aspects, as the initial conditions in the
wave-function and the validity of the Vakhitov-Kolokolov criterion.
For the initial conditions we considered a chirp parameter, which is
related to the initial focusing (defocusing) of the cloud. 
A non-zero initial chirp $b_0$ introduces oscillations in the
condensate near the minimum of the energy, such that it can lead a
previously stable system to collapse, when $b_0$ reaches a maximum limit
$b_{0,m}$, that corresponds to the energy difference between a minimum and
a maximum of the total potential energy. 
So, in the presence of a fixed value of the chirp, the number
of atoms of a stable condensate with attractive interaction is smaller
than the corresponding value when the chirp is zero.
In case that the potential has two minima (when cubic and quintic
terms are present in the NLSE), related to two phases of the 
condensate, a bifurcation phenomenon is predicted in the dependence of
the value of the initial chirp. A maximum initial chirp can affect the
system in such a way that  the oscillations can switch from the gas phase
to the liquid phase and vice-versa.

The present study was performed by using exact numerical
solutions of partial differential equation, as well as by a
corresponding variational approach. The analytical predictions based
on the time-dependent VA, using Gaussian ansatz, are qualitatively
confirmed by the exact  time-dependent numerical simulations.
We would like to point out the advantages of the Gaussian VA when
comparing with the moments method, which encounters some 
difficulties in deriving the collapse conditions even in the pure
attractive two-body case, for some classes of the initial data (the 
sign of the time dependent function in the equation for the
mean-square-radius, $\langle r^2\rangle$, is not defined \cite{Wadati}).

The VA starts to deviate from the exact results when the system is close
to the collapse conditions and particularly for the unstable solutions
(maxima) of the total energy. 
For this regions, where the Gaussian VA fails, one should improve
the ansatz or take the results as a qualitative picture to guide the exact
numerical calculations. 
In this respect, one should note that, for the NLSE without the trap
potential, it is well known that the time-dependent variational approach
fails to describe the region near the collapse~\cite{BergeR,ZakharovR}.

Considering that in this work we have shown the relevance of an
initial chirp parameter to study the stability of the condensate,
we should observe that a good estimate for such parameter relies in
a good approach to determining both minimum (stable) and maximum
(unstable) solutions for the total energy. The Gaussian VA has a
good agreement (quantitatively) with exact results for the minima, but can
only give a qualitative description for the maxima. Still we can 
observe that the VA calculations of the chirp parameter, when compared
with exact numerical calculations, are in reasonable agreement.

We also would like to point out that the present approach can be extended
to study the stability of chirped laser beams in inhomogeneous three
dimensional media with Kerr nonlinearity.

Finally, the main results of the present work are:
(i) A chirp parameter in the wave-function is shown to be useful
to study the initial conditions for a stable condensate to remain 
stable. 
(ii) we show that the VK criterion cannot be extended to
cases of harmonic trapped BEC when the nonlinear two and three-body terms
have opposite signs;
(iii) if an atomic system has repulsive two-body interaction, the
collapse is possible if the effective three-body interaction is negative.
In this perspective, one can use the observed critical number of atoms in
order to determine the corresponding three-body parameter.

\section*{Acknowledgments}
We are grateful to Funda\c{c}\~ao de Amparo \`a Pesquisa do
Estado de S\~ao Paulo (FAPESP-Brazil) for partial financial support. 
LT and TF also acknowledge partial support from Conselho Nacional de 
Desenvolvimento Cient\'\i fico e Tecnol\'ogico.

\end{multicols}
\end{document}